\documentclass[%
reprint,
amsmath,amssymb,
aps,
pra,
floatfix,
]{revtex4-1}

\usepackage{graphicx}
\usepackage{dcolumn}
\usepackage{bm}
\usepackage{color}
\usepackage{soul}

\usepackage{float} 

\begin{document}

\title{New opportunities for ultrafast and highly enantio-sensitive imaging and control of chiral nuclear dynamics: towards enantio-selective attochemistry}

\author{David Ayuso$^{1,2}$} \email{d.ayuso@imperial.ac.uk}

\affiliation{$^1$Department of Physics, Imperial College London, United Kingdom}
\affiliation{$^2$Max-Born-Institut, Berlin, Germany}

\raggedbottom

\begin{abstract}

The recently introduced \emph{synthetic} chiral light [D. Ayuso et al, \emph{Nat. Photon.} \textbf{13}, 866-871 (2019)] has opened up new opportunities for ultrafast and highly efficient imaging and control of chiral matter.
Here we show that the giant enantio-sensitivity enabled by such light could be exploited to probe chiral nuclear rearrangements during chemical reactions in an highly enantio-sensitive manner.
Using a state-of-the-art implementation of time-dependent density functional theory, we explore how the nonlinear response of the prototypical chiral molecule H$_2$O$_2$ changes as a function of its dihedral angle, which defines its handedness.
The macroscopic intensity emitted from randomly oriented molecules at even harmonic frequencies (of the fundamental) depends strongly on this nuclear coordinate.
Because of the ultrafast nature of such nonlinear interactions, the direct mapping between chiral dichroism and nuclear geometry provides a way to probe chiral nuclear dynamics at their natural time scales.
Our work paves the way for ultrafast and highly efficient imaging of enantio-sensitive dynamics in more complex chiral systems, including biologically relevant molecules.

\end{abstract}

\maketitle

The 21st century has witnessed major advances 
in the emerging field of \emph{attochemistry} \cite{Nisoli2019}, the goal of which is to visualize and to control the motion of electrons and nuclei during chemical reactions. 
These include the observation of purely electron motion in atoms \cite{Hentschel2001Nat,Drescher2002Nat,Uiberacker2007Nat,Ott2014Nat},
molecules \cite{Martin2007Sci,Smirnova2009Nat,Sansone2010Nat,Woerner2010,Calegari2014,Peng2019NatRevPhys}
and solids \cite{Cavalieri2007Nat,Schultze2014Sci,Luu2015Nat,Tao2016Sci,Uzan2020NatPhot},
as well as of highly correlated dynamics of electrons and nuclei in biologically relevant systems \cite{Polli2010Nat},
with unprecedented temporal resolution.
However, despite these groundbreaking achievements, distinguishing ultrafast left- and right-handed chiral dynamics is still challenging \cite{Oppermann2019JPCL,Schmid2019JPCL}.

Chirality plays key roles in a number of scientific areas, from particle physics to pharmacy to astronomy.
In general, an object is chiral when it cannot be superimposed to its mirror image.
Chiral molecules appear in pairs of left- and right-handed enantiomers.
Distinguishing them is vital, e.g. because they can present different biological activity \cite{Singh2010}.
To this goal, we can to make them interact with another chiral object, such as chiral light, and measure their enantio-sensitive response.

Traditional optical methods for chiral recognition include photoabsorption circular dichroism (CD) and optical rotation (OR).
Both techniques rely on weak chiro-optical effects, which arise beyond the electric-dipole approximation, posing major challenges for time-resolved measurements of ultrafast chiral dynamics \cite{Oppermann2019JPCL}.
Developing highly efficient all-optical approaches capable of tracking ultrafast chiral dynamics without relying on weak magnetic interactions is yet an important unsolved challenge.

More sophisticated methods bypass the need of relying on non-electric-dipole interactions by analyzing enantio-sensitive \emph{vectorial} observables \cite{OrdonezPRA2018}.
In photo-electron circular dichroism (PECD), this observable is the direction of the photoelectron current upon ionization with circularly polarized light
\cite{Ritchie1976PRA,Powis2000JCP,Bowering2001PRL,Garcia2003JCP,Lux2012Angew,Stefan2013JCP,Garcia2013NatComm,Janssen2014PCCP,Lux2015ChemPhysChem,Kastner2016CPC,Comby2016JPCL,Beaulieu2016FD,Beaulieu2017Science,Comby2018NatComm,Beaulieu2018NatPhys},
where the forward-backwards asymmetry can reach values on the order of $10\%$.
Such a strong enantio-sensitivity has enabled the determination of the enantiomeric excess of mixtures with accuracy below $1\%$ \cite{Kastner2016CPC,Comby2018NatComm}
as well as ultrafast time-resolved measurements of chiral dynamics \cite{Comby2016JPCL,Beaulieu2016FD}.
The recent application or tailored 2-colour fields to PECD has lead to the generation of highly enantio-sensitive \emph{tensorial} observables \cite{Demekhin2018PRL,Goetz2019PRL,RozenPRX2019}.
However, the total intensity signal (i.e. the total number of emitted photoelectrons) is barely enantio-sensitive, and these methods require angularly resolved measurements of photoelectrons.

Using microwave radiation, Patterson and co-workers \cite{Patterson2013} pioneered the development of a highly efficient way of enantio-discrimination, within the electric-dipole approximation.
More recent achievements include the generation of enantio-sensitive populations of rotational states using fields with three non-coplanar frequency components \cite{Eibenberger2017PRL,Shubert2016,Perez2017},
as well as the development chiral optical centrifuges \cite{Yachmenev2019PRL,Milner2019PRL} for enantio-separation.
These approaches enable efficient control over the \emph{rotational} degrees of freedom of chiral molecules in a highly enantio-sensitive way. 
Achieving full control over chemical reactions using light --the ultimate goal of attochemistry-- requires acting on the \emph{electronic} degrees of freedom, on the attosecond and femtosecond time scales.

Synthetic chiral light \cite{Ayuso2019NatPhot} enables all-optical imaging of chiral matter and ultrafast chiral dynamics at the level of electrons, with extremely high enantio-sensitivity. 
Such light is \emph{locally} chiral, i.e. within the electric-dipole approximation: the tip of the electric field vector draws a chiral, three-dimensional Lissajous figure in time.
The enantio-sensitive response of chiral matter arises within the electric-dipole approximation, and therefore is orders-of-magnitude stronger than in traditional chiro-optical methods.
We have recently shown that the application of synthetic chiral light \cite{Ayuso2019NatPhot} to chiral high-harmonic generation (HHG) can dramatically enhance the enantio-sensitivity of this emerging technique,
which so far had relied on the interplay of the chiral molecules with the magnetic component of the light wave \cite{Cireasa2015NatPhys,Ayuso2018JPB,Ayuso2018JPB_model,Harada2018PRA,Baykusheva2018PRX}.
We have also introduced several all-optical schemes based on purely electric-dipole interactions where the molecular handedness is encoded in the polarization \cite{Neufeld2019PRX,Ayuso2021Optica} or in the direction of emission \cite{Ayuso2021NatComm} of the harmonic light.

The giant enantio-sensitivity enabled by synthetic chiral light opens up promising opportunities for ultrafast imaging and control of chiral matter.
Can we exploit it to monitor the nuclear rearrangements occurring during enantio-sensitive chemical reactions, in real time, and with high enantio-sensitivity?
We aim to provide (positive) answer this question.

This paper is organized as follows.
First, we review the key aspects of synthetic chiral light, introduced in \cite{Ayuso2019NatPhot}. 
Second, we describe the method that we used to compute ultrafast nonlinear dynamics in the prototypical chiral molecule H$_2$O$_2$, which is a convenient model to test our ideas.
Third, we show (numerically) how the enantio-sensitive response of this molecule depends strongly on its nuclear coordinates, opening new opportunities for ultrafast imaging of chemical dynamics.
We conclude by discussing these opportunities.

\section{Synthetic chiral light}

The ``recipe'' for creating synthetic chiral light has two main ``ingredients'' \cite{Ayuso2019NatPhot}.
First, we need at least two phase-locked frequencies.
Phase locking is key because the handedness of synthetic chiral light is sensitive to the multi-colour phase delays.
Second, we require a relatively strong longitudinal component, i.e. an electric field component in the propagation direction, which is absent in standard light, but appears naturally in non-collinear configurations or tightly focused beams \cite{Bliokh2015}.
This longitudinal component allows the Lissajous curve of the field to be 3D and chiral.

\begin{figure}
\centering
\includegraphics[width=\linewidth, keepaspectratio=true]{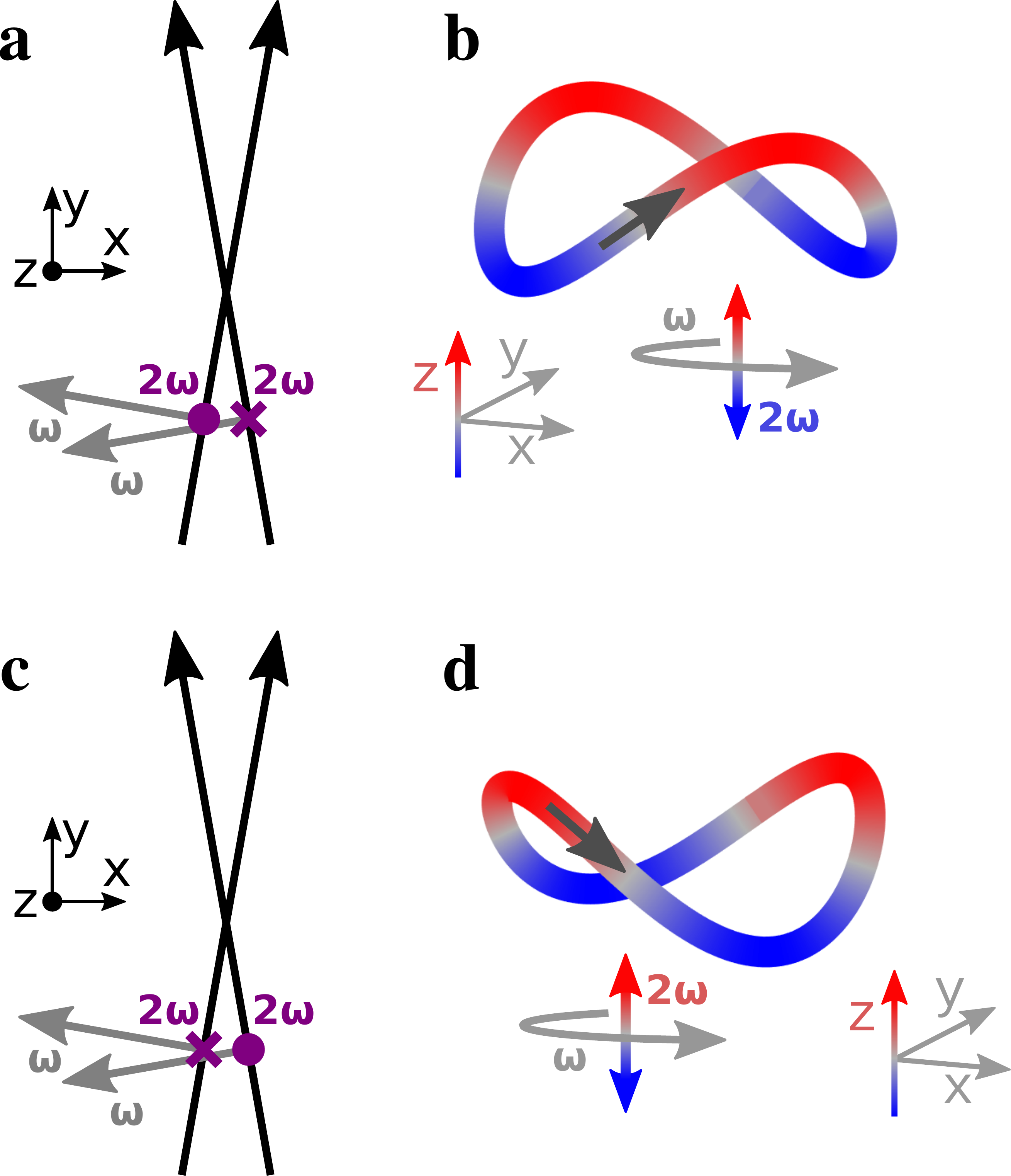}
\caption{
\textbf{Synthetic chiral light}.
\textbf{a,} A \emph{locally} and \emph{globally} chiral field can be created using two laser beams that propagate noncollinearly and carry two colours: a fundamental $\omega$ frequency and its second harmonic, with opposite $\omega-2\omega$ phase delay in both beams.
\textbf{b,} Lissajous figure drawn by the tip of the total electric field vector in the overlap region between the two beams, i.e.  $\phi_{\omega,2\omega}^{(2)}=\phi_{\omega,2\omega}^{(1)}+\pi$.
\textbf{c,d,} Changing the $\omega-2\omega$ phase delay in the two beams by $\pi$, e.g. by reflecting the optical setup on the $xy$ plane (\textbf{c}), reverses the handedness of the chiral Lissajous figure (\textbf{d}).
}
\label{fig_SCL}
\end{figure}

Here we use the original implementation of synthetic chiral light, introduced in \cite{Ayuso2019NatPhot}.
The proposed optical setup is depicted in Fig. \ref{fig_SCL}.
It requires two laser beams that propagate noncollinearly in the $xy$ plane, creating small angles $\pm\alpha$ with respect to the $y$ direction, i.e.
\begin{align}
\textbf{k}_1 &=   k \sin(\alpha) \hat{\textbf{x}} + k \cos(\alpha) \hat{\textbf{y}} \\
\textbf{k}_2 &= - k \sin(\alpha) \hat{\textbf{x}} + k \cos(\alpha) \hat{\textbf{y}}
\end{align}
where $k=2\pi/\lambda$, with $\lambda$ being the fundamental wavelenght.
Both beams carry two orthogonally polarized colours: the fundamental frequency $\omega$, polarized in the $xy$ propagation plane, and its $z$-polarized second harmonic:
\begin{align}
\textbf{F}_n^{ \omega} &= F_0^{ \omega} e^{-\rho^2/w^2} \cos( \textbf{k}_n\cdot\textbf{r} -  \omega t ) \textbf{e}_n \\
\textbf{F}_n^{2\omega} &= F_0^{2\omega} e^{-\rho^2/w^2} \cos(2\textbf{k}_n\cdot\textbf{r} - 2\omega t - \phi_{2\omega}^{(n)} ) \hat{\textbf{z}}
\end{align}
where $\rho$ is the distance to the beam's axis, $w$ is the beam's waist, and the polarization vectors of the $\omega$-field component are defined as:
\begin{align}
\textbf{e}_1 &= \cos(\alpha) \hat{\textbf{x}} - \sin(\alpha) \hat{\textbf{y}} \\
\textbf{e}_2 &= \cos(\alpha) \hat{\textbf{x}} + \sin(\alpha) \hat{\textbf{y}}
\end{align}
In the overlap region, the total electric field can be written as
\begin{equation}\label{eq_field}
\textbf{F} = F_{\omega} \big[ \cos(\omega t) \hat{\bold{x}} + \varepsilon \sin(\omega t) \hat{\bold{y}} \big] + F_{2\omega} \cos(2\omega t + \phi_{2\omega}) \hat{\bold{z}}
\end{equation}
where the $\omega-2\omega$ phase delay $\phi_{2\omega}$ is related to the $\omega-2\omega$ phase delays in the individual beams by
\begin{equation}
\phi_{2\omega} = \frac{\phi_{2\omega}^{(1)}+\phi_{2\omega}^{(2)}}{2}
\end{equation}
and
\begin{align}
F_{\omega}(x)  &= 2 F_0^{ \omega} e^{-\rho^2/w^2} \cos(\alpha) \cos( k_\alpha x) \\
F_{2\omega}(x) &= 2 F_0^{2\omega} e^{-\rho^2/w^2}              \cos(2k_\alpha x + \phi_{-}) \\
\varepsilon(x) &= - \tan(\alpha) \tan(k_\alpha x)
\end{align}
where $k_\alpha = k \sin(\alpha)$, and we assume that the interaction region is sufficiently thin so $\rho$ barely changes.
Eq. \ref{eq_field} shows that the total electric field is elliptically polarized at frequency $\omega$ in the $xy$ plane, with the minor ellipticity component along $y$,
and has $z$-polarized $2\omega$ component, creating the chiral Lissajous figure shown in Fig. \ref{fig_SCL}b.

Locally chiral fields can drive strongly enantio-sensitive interactions at the single-molecule-response level, which are controlled and characterized via chiral correlation functions, see \cite{Ayuso2019NatPhot}.
However, imprinting this \emph{microscopic} response into the total \emph{macroscopic} signal, requires an additional condition: the field's handedness needs maintained in space, at least to certain extent over the interaction region.
That is, synthetic chiral light also needs to be \emph{globally} chiral.
In our setup, this condition is fulfilled if $\phi_{2\omega}^{(2)}=\phi_{2\omega}^{(1)}+\pi$.
Then, by changing these phase delays synchronously, we can tailor the shape of the chiral Lissajous figure in a way that its handedness is the same all over the interaction region.
In particular, if we change these phase delays by $\pi$ in both beams (Fig. \ref{fig_SCL}c), we reverse the handedness of the chiral Lissajous figure (Fig. \ref{fig_SCL}d).
Synthetic chiral light that is \emph{locally} and \emph{globally} chiral enables the highest possible degree of control over the enantio-sensitive chiral matter: quenching it in one molecular enantiomer while enhancing it in its mirror twin.

\begin{figure}
\centering
\includegraphics[width=\linewidth, keepaspectratio=true]{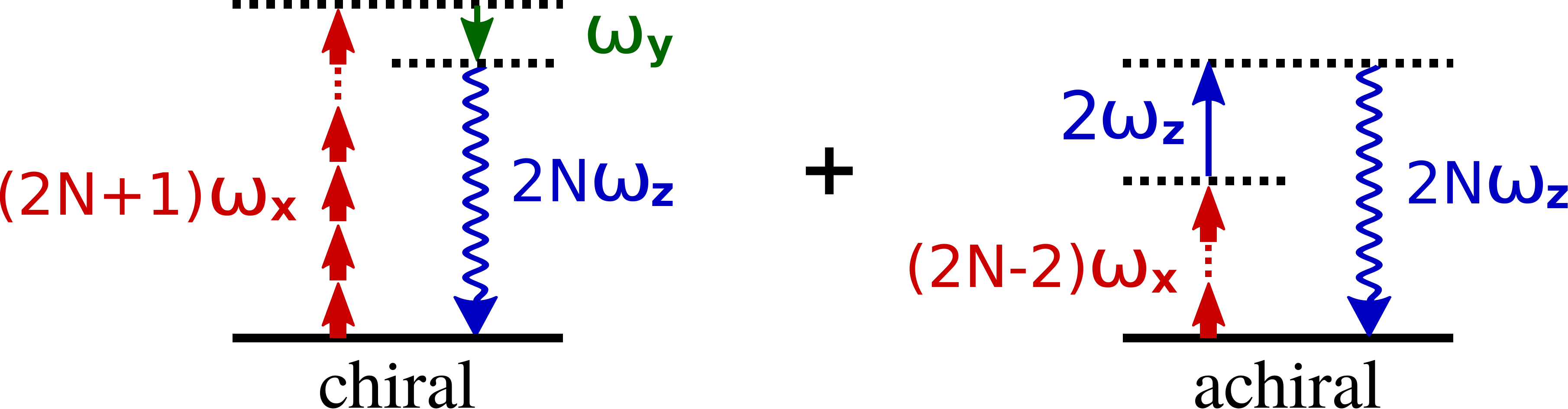}
\caption{\textbf{Multiphoton diagrams} describing the enantio-sensitive response of isotropic chiral matter to synthetic chiral light in the perturbative regime.
It relies on the interference between two contributions to light-induced polarization: a chiral contribution (left diagram) and an achiral contribution (right diagram), see main text.
}
\label{fig_diagram}
\end{figure}

The enantio-sensitive response of isotropic chiral matter to our locally chiral field relies on the interference of chiral and achiral contributions to light-induced polarization.
The multiphoton pathways describing these transitions in the perturbative regime are depicted in Fig. \ref{fig_diagram}.
The chiral contribution involves absorption of $2N+1$ photons from the (x-polarized) strong-field component and emission of $1$ photon into the (y-polarized) weak ellipticity component, 
leading to polarization at frequency $2N\omega$ along $z$.
This pathway is exclusive of chiral media, and the induced polarization is out of phase in media of opposite handedness.
The chiral contribution involves absorption of $2N-2$ photons from the strong-field component and absorption of $1$ photon from the z-polarized $2\omega$-field,
also leading to polarization at frequency $2N\omega$ along $z$.
As these two pathways lead to polarization at the same frequency and along the same axis, they interfere.
This interference can be controlled and quantified via chiral correlation functions, see \cite{Ayuso2019NatPhot}.

\section{Modeling the nonlinear response of H$_2$O$_2$ to synthetic chiral light}

\begin{figure}
\centering
\includegraphics[width=\linewidth, keepaspectratio=true]{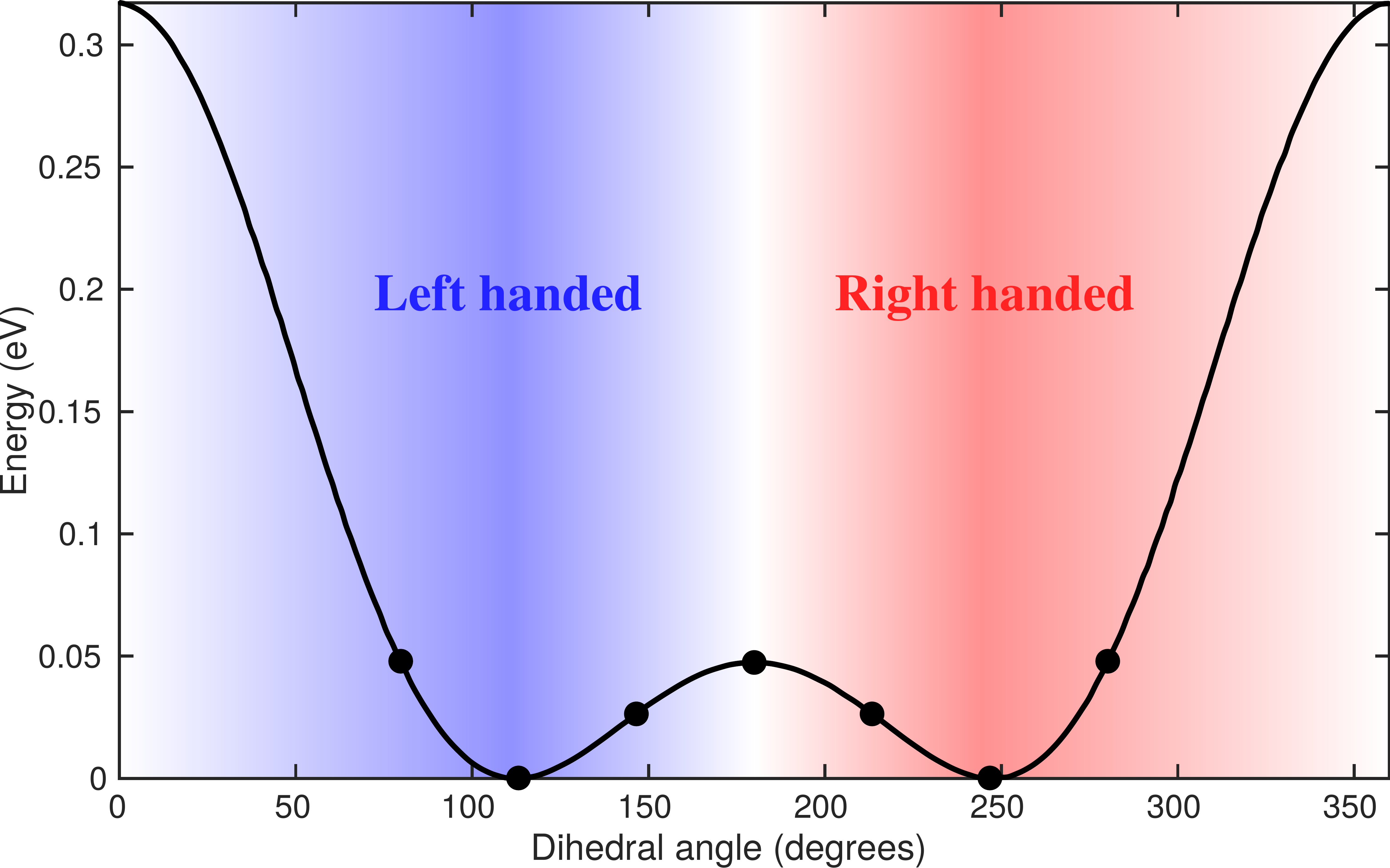}
\caption{\textbf{Hydrogen peroxide as a prototypical chiral molecule with reversible handedness}.
Potential energy curve of the electronic ground state of H$_2$O$_2$ as a function of the dihedral angle $\alpha$ calculated at the B3LYP theory level; the numerical values have been taken from Fig. 1 of \cite{Kuenzer2019}.
The points indicate the values of the dihedral angle that we have considered in this work to evaluate the ultrafast electronic response of the molecule to synthetic chiral light.
The molecule is achiral if $\alpha=0^\circ$ or $\alpha=180^\circ$, ``left-handed'' if $\alpha<180^\circ$, and ``right-handed'' if $\alpha>180^\circ$.
The nuclear configurations with dihedral angles $\alpha$ and $360-\alpha$ correspond to molecular enantiomers.
}
\label{fig_PEC}
\end{figure}

The hydrogen peroxide molecule constitutes a convenient ``toy model'' to test the feasibility of synthetic chiral light to monitor nuclear dynamics in real time.
The potential energy curve of the electronic ground state is shown in Fig. \ref{fig_PEC}.
It presents a double-well structure, with the two degenerate geometries associated with these wells being molecular enantiomers. 
Indeed, the molecular geometries with dihedral angles $\alpha$ and $360^\circ-\alpha$ are mirror twins.
However, since the energy barrier separating these two wells is small, of around $0.5$eV, the two enantiomers co-exist in standard conditions.
Still, from a computational perspective, the small size of this chiral molecule makes it an ideal candidate to test our ideas. 

We have modeled the interaction of several nuclear configurations of H$_2$O$_2$ with synthetic chiral light using the state-of-the-art implementation of time-dependent density functional theory (TDDFT) in Octopus \cite{Marques2003,Castro2006,Andrade2015,Tancogne2020}.
We used of the local-density approximation \cite{Dirac1930,Bloch1929,Perdew1981} to account for electronic exchange and correlation effects,
and the averaged-density self-interaction correction \cite{Legrand2002} to ensure an accurate description of the long-range part of the electronic wave function.
The 1s orbitals of the oxygen atoms were described via pseudo-potentials.
We used a uniform real-space grid in a spherical basis set of radius $R=30.7$ a.u., with spacing between neighboring points of $0.4$ a.u., and a complex absorbing potential with width $20$ a.u. and height $-0.2$ a.u.

To describe the physical situation of randomly oriented molecules, we run simulations for different molecular orientations.
The total polarization results from the coherent addition of the contribution from all possible orientations:
\begin{equation}
\mathbf{P}(N\omega) = \int d\Omega \int d\alpha \, \mathbf{P}_{\Omega,\alpha}(N\omega),
\end{equation}
where $N$ is the harmonic number and $\omega$ is the fundamental frequency.
We used a Lebedev quadrature of order 11 \cite{Lebedev1999} to integrate over the solid angle $\Omega$.
The laser field was rotated in the molecular frame in a way that the orientation of the strong-field component depended on $\Omega$ but not on $\alpha$, which allowed us to reach convergence using only 4 points in $\alpha$.

We have run numerical simulations for the left-handed configurations with dihedral angles $79^\circ$, $112^\circ$ and $146^\circ$ (see Fig. \ref{fig_PEC}), and for four values of the $\omega-2\omega$ phase delay: $0$, $\pi/2$, $\pi$ and $3\pi/2$.
The results for other phase delays and for the right-handed configurations were obtained via numerical interpolation and using symmetry considerations.

In the proposed implementation of synthetic chiral light (see Fig. \ref{fig_SCL}), the ($y$-polarized) ellipticity component and the ($z$-polarized) $2\omega$-field component are sufficiently weak so the response along $y$ and $z$ can be treated perturbately.
That is, the chiral response is essentially unaffected by presence of the $2\omega$ field, and the achiral is not modified by the weak ellipticity component.
To evaluate the single-molecule response in the interaction region, we also assumed that the nonlinear response along $x$ depends on the number of absorbed photons as $D_x \propto I^{N/2}$,
and run simulations for intensity $I=10\cdot10^{14}$ W$\cdot$cm$^{-2}$, with fundamental wavelength $\lambda=400$ nm ($\omega=0.114$ a.u.).

We used a sine-squared flat-top envelope of 8 laser cycles ($10.7$ fs): 2 cycles to rise up and to go down at the beginning and end of the simulation (sine-squared function), and 4 cycles of constant intensity in the middle.
To reduce the numerical noise in the harmonic spectrum, we applied a Gaussian filter to the induced polarization in the time domain.
This filter had a full-width at have maximum of $2.7$ fs and was centered at the center of the envelope.

The far-field image was evaluated using the Fraunhofer diffraction equation, as in our previous works \cite{Ayuso2019NatPhot,Neufeld2019PRX,Ayuso2021NatComm,Ayuso2021Optica}.

\section{Numerical results}

We now present numerical results of the harmonic response driven by the optical setup presented in Fig. \ref{fig_SCL} in hydrogen peroxide.
Symmetry dictates that the enantio-sensitive signal is completely background free, separated from the non-enantio-sensitive response in frequency, polarization and space \cite{Ayuso2019NatPhot}.
The non-enantio-sensitive signal appears at odd harmonic frequencies and is polarized along x, whereas the enantio-sensitive response arises at even harmonic frequencies, and is polarized along z.
They are also emitted in different directions, as discussed in our previous work \cite{Ayuso2019NatPhot}.

\begin{figure}
\centering
\includegraphics[width=\linewidth, keepaspectratio=true]{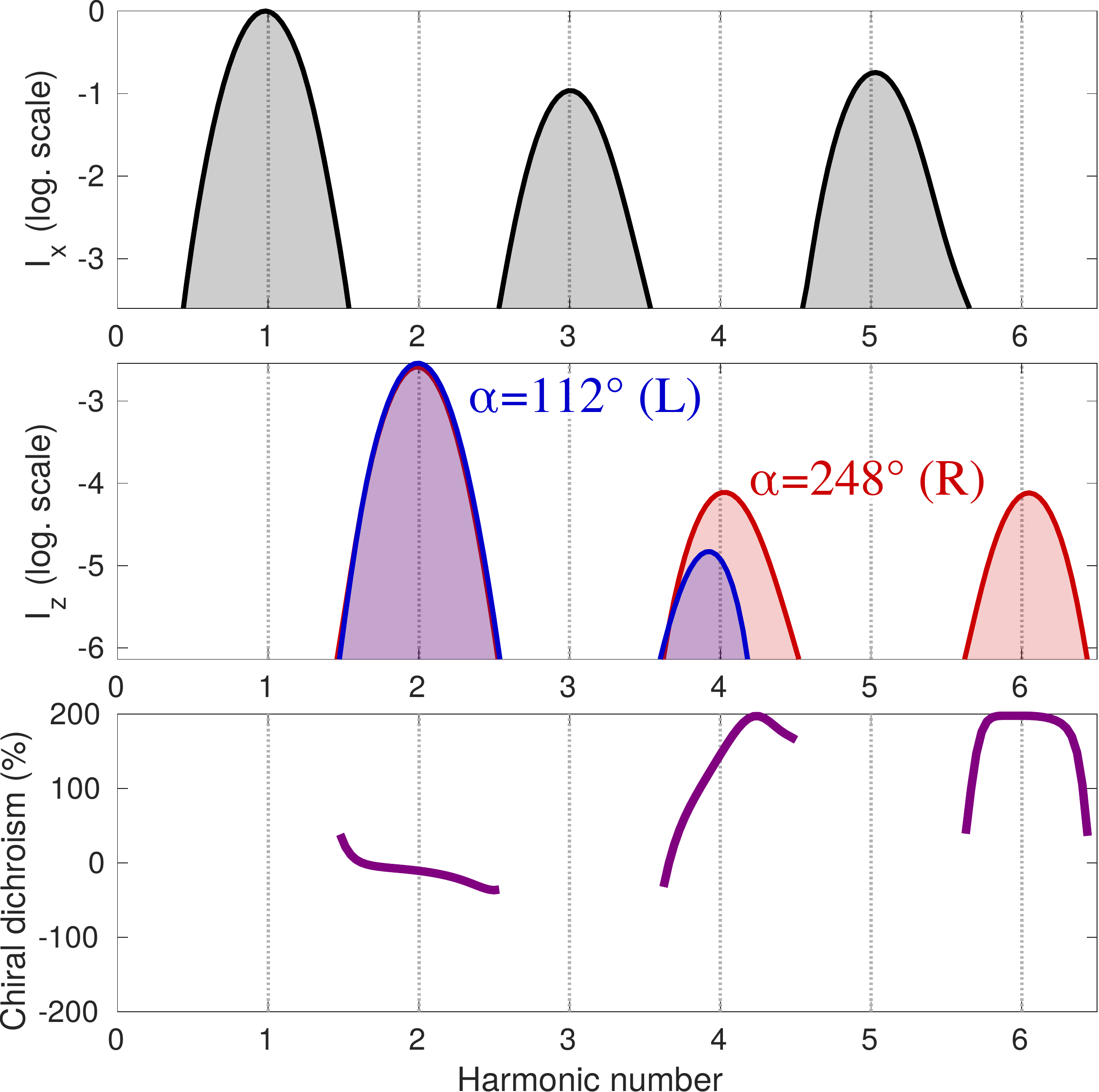}
\caption{\textbf{Macroscopic nonlinear response} of the equilibrium geometries of H$_2$O$_2$ to synthetic chiral light.
\textbf{a,} Non-enantio-sensitive x-polarized harmonic intensity emitted from the left-handed ($\alpha=112^\circ$, see Fig. \ref{fig_PEC}) and right-handed ($\alpha=248^\circ$) equilibrium geometries.
\textbf{b,} Enantio-sensitive z-polarized harmonic intensity emitted from the left-handed ($\alpha=112^\circ$, blue) and right-handed ($\alpha=248^\circ$, right) equilibrium geometries.
\textbf{c,} Chiral dichroism, see Eq. \ref{eq_CD}.
}
\label{fig_HHS}
\end{figure}

Fig. \ref{fig_HHS} shows the far-field harmonic intensity emitted by the left- and right-handed equilibrium configurations of H$_2$O$_2$.
We considered two laser beams with fundamental wavelength $\lambda=400$nm and intensity $I_{\omega}=2.5\cdot10^{13}$ W$\cdot$cm$^{-2}$, so the total intensity in the overlap region reaches $10^{14}$ W$\cdot$cm$^{-2}$.
The angle between the propagation directions and the $y$ axes was set to $\alpha=\pm10^{\circ}$, the relative field strength of the $2\omega$-field component was set to $\sqrt{I_{2\omega}/I_{\omega}}=0.0125$ in both beams,
and the relative phase delays to $\phi_{2\omega}^{(1)}=0.1\pi$ and $\phi_{2\omega}^{(1)}=1.1\pi$, so the 2-colour phase delay of the locally chiral field was $\phi_{2\omega}=0.6\pi$ (Eq. \ref{eq_field}) all over the interaction region.

The non-enantio-sensitive (Fig. \ref{fig_HHS}a) and the enantio-sensitive (Fig. \ref{fig_HHS}b) harmonic signals appear at different harmonic orders and are orthogonally polarized, as above discussed, making the experimental detection easier.
Fig. \ref{fig_HHS}b shows that the enantio-sensitive (z-polarized) intensity emitted from the left- and right-handed enantiomers of H$_2$O$_2$ can be dramatically different.
Here, we have tuned the relative amplitude and phase of the $2\omega$-field component to quench emission of harmonic 6 (H6) in the left-handed enantiomer, while maximizing it in the right-handed enantiomer.

To quantify the degree of enantio-sensitivity, we use the standard definition of chiral dichroism:
\begin{equation}\label{eq_CD}
CD = 2 \frac{I_L-I_R}{I_L-I_R}
\end{equation}
For the case of H$_2$O$_2$, as the nuclear configurations with dihedral angles $\alpha$ and $360-\alpha$ are molecular enantiomers, we can easily connect the chiral dichroism to the molecular geometry via
\begin{align}
I_L &= I(\alpha) \\
I_R &= I(360^\circ-\alpha)
\end{align}
where $0<\alpha<180^\circ$.
In the ground-state configurations ($\alpha=112^\circ$, see Fig. \ref{fig_PEC}), the chiral dichroism, shown in Fig. \ref{fig_HHS}c, reaches its maximum value: $200\%$.

An alternative expression of chiral dichroism can be defined using the intensity ratio between an enantio-sensitive even ($z$-polarized) harmonic and an odd ($x$-polarized) harmonic,
\begin{equation}\label{eq_CD2}
CD' = 2 \frac{I_L^{2N}/I_L^{2N+1}-I_R^{2N}/I_R^{2N+1}}{I_L^{2N}/I_L^{2N+1}+I_R^{2N}/I_R^{2N+1}},
\end{equation}
rather than absolute intensities (as in Eq. \ref{eq_CD}).
This alternative definition could reduce errors associated with experimental fluctuations in the laser intensity and sample densities.
From a theoretical perspective, both definitions are equivalent, as the reference ($x$-polarized harmonics) are not enantio-sensitive, i.e. $I_L^{2N+1}=I_R^{2N+1}$, and thus $CD=CD'$.

\begin{figure}
\centering
\includegraphics[width=\linewidth, keepaspectratio=true]{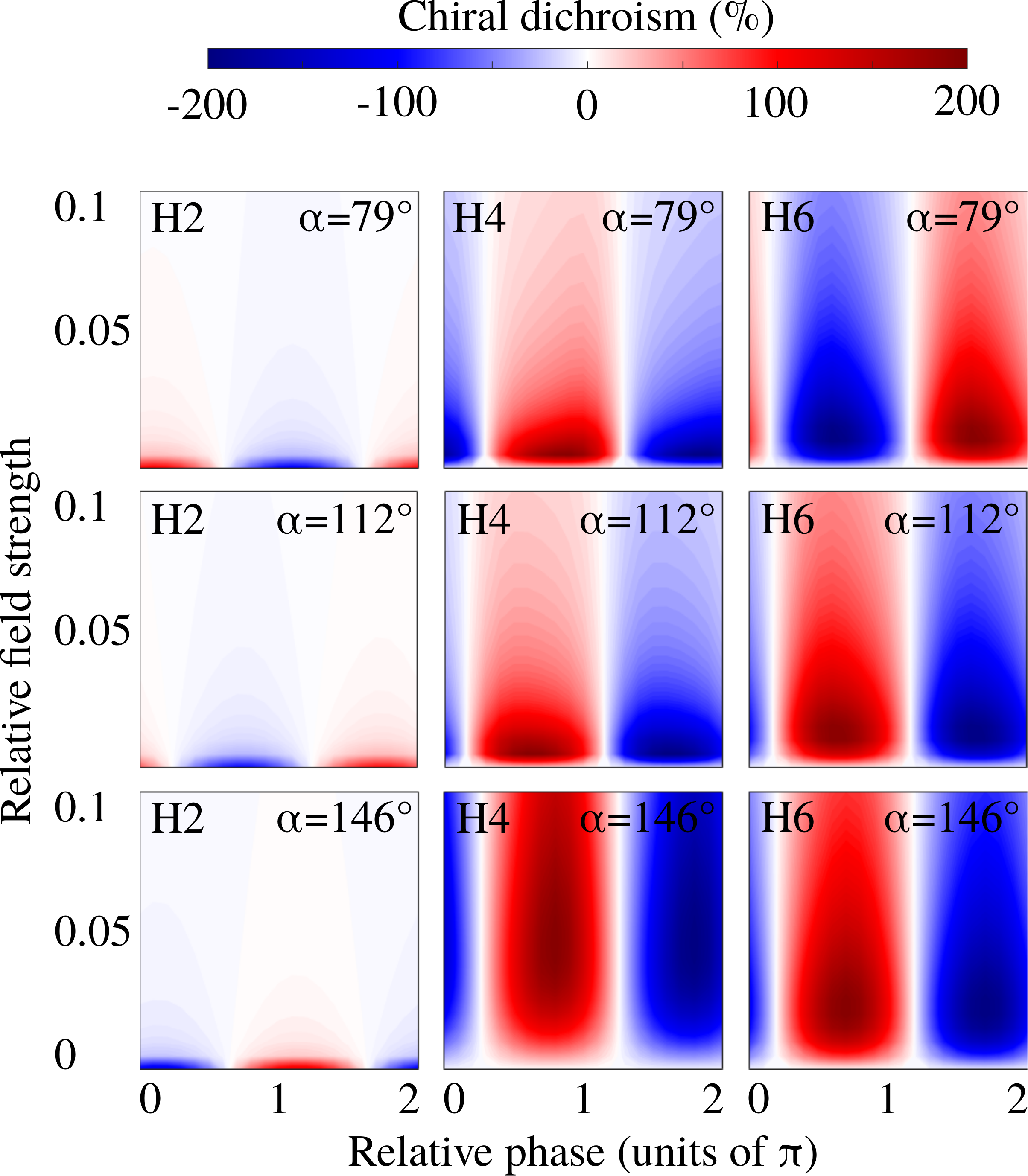}
\caption{\textbf{Chiral dichroism} in the enantio-sensitive $z$-polarized harmonic intensity emitted from randomly oriented H$_2$O$_2$ molecules (see Eq. \ref{eq_CD})
as a function the $\omega-2\omega$ phase delay (\textbf{a}) and relative $2\omega$-field strength,
for harmonic numbers 2 (left panels), 4 (central panels) and 6 (right panels),
and for dihedral angles $79^\circ$ (upper rows), $112^\circ$ (central rows), $146^\circ$ (lower rows).
}
\label{fig_CD}
\end{figure}

The enantio-sensitive response to synthetic chiral light relies on the interference between two contributions to light-induced polarization, as above discussed.
The chiral contribution is out of phase in media of opposite handedness and is not affected by the $2\omega$-field component (Fig. \ref{fig_diagram}, left).
The achiral contribution is identical in opposite enantiomers, and its intensity and phase are fully controlled by the relative strength and phase $2\omega$-field component  (Fig. \ref{fig_diagram}, right).
By adjusting these parameters, we can achieve perfect constructive interference in one enantiomer and perfect destructive interference the other.
As we show in the following, this allows us to reach reach the limits of chiral dichroism ($\pm200\%$) in any harmonic number and for any nuclear configuration.


Fig. \ref{fig_CD} shows the chiral dichroism for H2, H4 and H6, and for dihedral angles $79^\circ$, $112^\circ$ and $146^\circ$,
as a function of the relative strength and phase of the $2\omega$-field component.
As already anticipated, we reach the limits of chiral dichroism ($\pm200\%$) in all cases.
Note that changing $\phi_{\omega,2\omega}$ by $\pi$ reverses the field's handedness.
As a result, the values of $\phi_{\omega,2\omega}$ that maximize the response of the left-handed enantiomer ($CD=200\%$) and the right-handed enantiomer ($CD=-200\%$) are always shifted by $\pi$.
Furthermore, in the perturbative regime, the phase of the achiral response depends linearly on $\phi_{\omega,2\omega}$.
Thus, the nodal lines in Fig. \ref{fig_CD}, corresponding to $CD=0$, are shifted by $\pi/2$ with respect to the maximum and the minimum values of $CD$.
The specific features of these two-dimensional spectrograms record the nuclear configuration of the molecule.

\begin{figure}
\centering
\includegraphics[width=\linewidth, keepaspectratio=true]{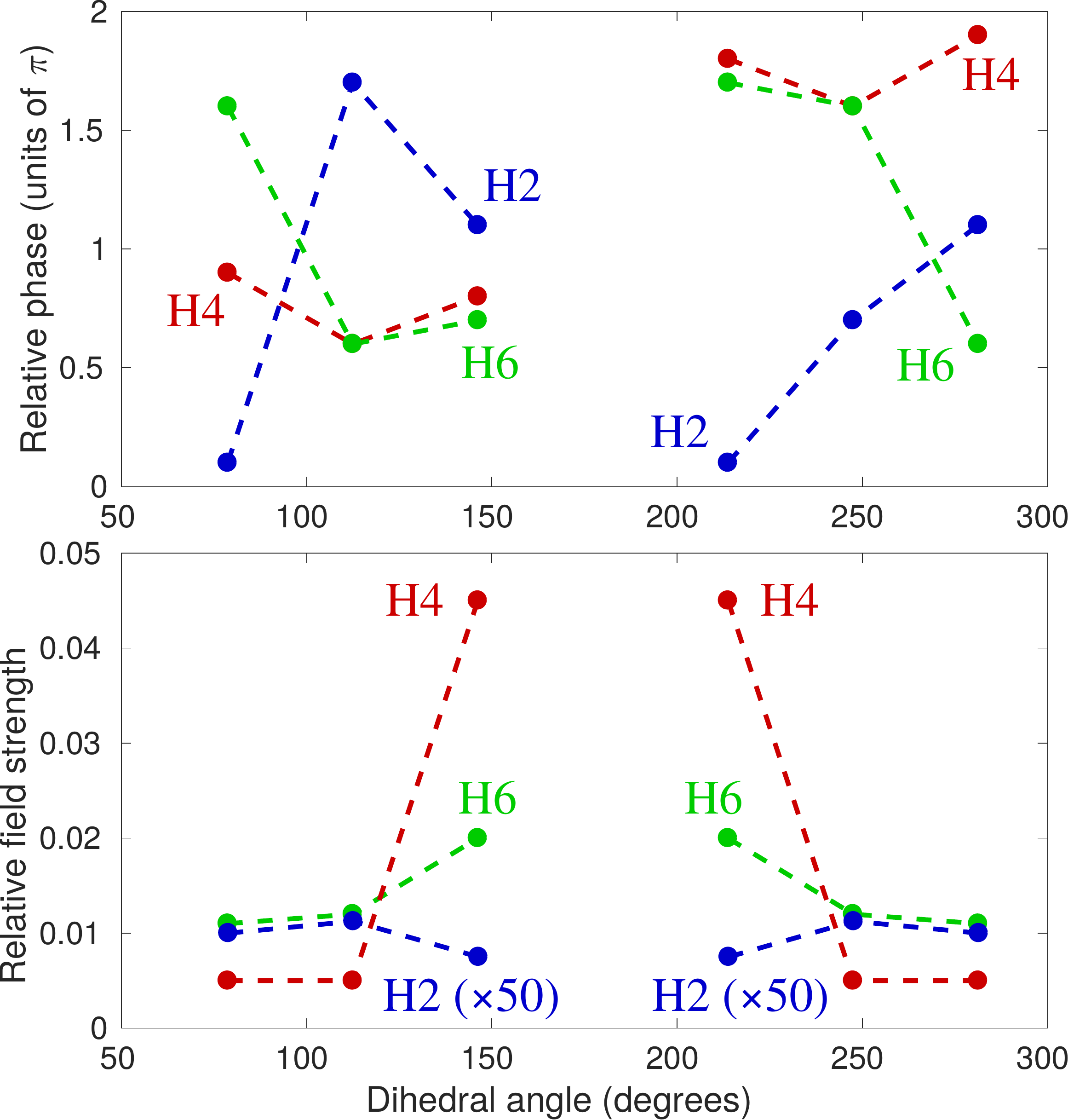}
\caption{\textbf{Maximizing the enantio-sensitive response of hydrogen peroxide}.
Values of the $\omega-2\omega$ phase delay (\textbf{a}) and of the relative $2\omega$-field strength, i.e. $F_{2\omega}/F_{\omega}$ (\textbf{b}) which maximize the ($z$-polarized) enantio-sensitive response of H$_2$O$_2$ as a function of the dihedral angle,
for harmonic numbers 2 (blue), 4 (red) and 6 (green).
}
\label{fig_optim}
\end{figure}

The optimum values of relative field strength $\sqrt{I_{2\omega}/I_{\omega}}$ and relative phase $\phi_{2\omega}$ which maximize the enantio-sensitive response of the left-handed enantiomer ($CD=200\%$ in Fig. \ref{fig_CD})
are presented in Fig. \ref{fig_optim}, as a function of the dihedral angle.
They are different for each harmonic number, as they are associated with different multiphoton processes.
Because the molecular susceptibilities change with the dihedral angle, these optimum parameters are also different for different molecular geometries.

The strong modulation of both spectroscopic parameters with the dihedral angle shows the feasibility of synthetic chiral light to probe the nuclear configuration of the molecule in a highly enantio-sensitive manner.
Because of the ultrafast nature of such nonlinear interactions, synthetic chiral light may allow us to probe chemical reactions in real time.

\section{Conclusions}

Monitoring and controlling ultrafast chemical processes at their natural time scales are the ultimate goals of attochemistry. 
These goals are particularly challenging when dealing with chiral molecules, as the weakness of non-electric-dipole interactions makes the response of left- and right-handed enantiomers to conventional light fields virtually identical.
Synthetic chiral light, which is chiral already in the electric dipole approximation, overcomes this fundamental limitation.
It creates a new way of imaging and controlling the electronic clouds in chiral molecules with extremely high enantio-sensitivity, bringing new opportunities to this emerging discipline.

Here we have shown that the ultrafast electronic response of the prototypical chiral molecule H$_2$O$_2$ is highly sensitive to its nuclear geometry. 
Despite the simplicity of the molecular system, our numerical results validate the feasibility of synthetic chiral light as a tool for imaging enantio-sensitive chemical reactions in real time,
which are ubiquitous in chemistry and of special relevance in biology.
Further research should address more complex chiral systems, where synthetic chiral light may allow us to image and control the correlated interplay between electronic and nuclear degrees of freedom, which becomes particularly important in the vicinity of conical intersections.

Other interesting opportunities include the possibility of initiating highly enantio-sensitive dynamics in mixtures of left- and right-handed molecules with synthetic chiral light.
It may allow us to initiate photo-chemical reactions in a highly enantio-sensitive way.

\section*{Acknowledgements}
I am extremely grateful to Olga Smirnova and Misha Ivanov for their enormous support and guidance, for pushing me in the right direction, for highly stimulating discussions, and for being a constant source of inspiration.
I also acknowledge very stimulating discussions with Andr\'es F. Ordo\~nez, expert advice from Oriana Brea in computing the geometry optimizations, and funding from the Royal Society URF$\backslash$R1$\backslash$201333 and the Deutsche Forschungsgemeinschaft SPP 1840 SM 292/5-2
\bibliography{Bibliography}

\end{document}